\documentclass[10pt]{IEEEtran}
\usepackage{amsmath,amssymb}
\usepackage{latexsym,setspace,graphicx,url,tabularx,wrapfig,multirow,epsfig,enumerate}
\usepackage{mathptmx}
\usepackage{xspace}
\usepackage{paralist}
\usepackage{slashbox}
\usepackage{ifpdf}
\usepackage{color}

\usepackage{authblk}
\usepackage{floatrow}

\usepackage{graphicx}
\usepackage{caption}
\usepackage{subcaption}
\usepackage{hhline}
\usepackage{threeparttable}

\usepackage{wrapfig}
\ifpdf
    \DeclareGraphicsExtensions{.pdf,.png,.jpg,.mps}
\else
    \DeclareGraphicsExtensions{.eps,.ps}
\fi

\newcommand{\Xomit}[1]{}
\newcommand{\alglight}{Sample\&HH }
\newcommand{\algfull}{Sample\&Pick }
\newcommand{\algsnh}{Sample\&Hold }
\newcommand{\alglightNoSp}{Sample\&HH}
\newcommand{\algfullNoSp}{Sample\&Pick}
\newcommand{\algsnhNoSp}{Sample\&Hold}

\newtheorem{theorem}{Theorem}[section]

\newtheorem{proof-sketch}{Proof Sketch}
\newtheorem{definition}[theorem]{Definition}

\usepackage[longend,boxruled]{algorithm2e}
\DontPrintSemicolon
\SetKwInOut{Input}{input}
\SetKwInOut{Output}{output}

\begin{document}

\clubpenalty=10000
\widowpenalty = 10000


\title{Detecting Heavy Flows in the SDN Match and Action Model}

\author[1]{Yehuda Afek}
\author[2]{Anat Bremler-Barr}
\author[1]{Shir Landau Feibish}
\author[1]{Liron Schiff}
\affil[1]{Blavatnik School of Computer Science, Tel-Aviv University, Israel}
\affil[2]{Computer Science Dept., Interdisciplinary Center, Herzliya, Israel}
\affil[ ]{\textit{afek@cs.tau.ac.il, bremler@idc.ac.il, shirl11@post.tau.ac.il, schiffli@post.tau.ac.il}}

\maketitle
\thispagestyle{empty}

{\let\thefootnote\relax\footnote{This research was supported by European Research Council (ERC) Starting Grant no. $259085$, and the Neptune Consortium, administered by the Office of the Chief Scientist of the Israeli ministry of Industry, Trade, and Labor, and the Ministry of Science and Technology, Israel.}
}


\begin{abstract}

Efficient algorithms and techniques to detect and identify large flows in a high throughput traffic stream in the SDN match-and-action model are presented. 
This is in contrast to previous work that either deviated from the match and action model by requiring additional switch level capabilities or did not exploit the SDN data plane.
Our construction has two parts; (a) how to sample in an SDN match and action model, (b) how to detect large flows efficiently and in a scalable way, in the SDN model.

Our large flow detection methods provide high accuracy and present a good and practical tradeoff between switch - controller traffic, and the number of entries required in the switch flow table. Based on
different parameters, we differentiate between heavy flows,
elephant flows and bulky flows and present efficient algorithms to
detect flows of the different types.

Additionally, as part of our heavy flow detection scheme, we present sampling methods to sample packets with arbitrary probability $p$ per packet or per byte that traverses an SDN switch.

Finally, we show how our algorithms can be adapted to a distributed monitoring SDN setting with multiple switches, and easily scale with the number of monitoring switches.

\end{abstract}

\section{Introduction}
\label{Section:Introduction}
Heavy flow detection in traffic remains one of the fundamental capabilities required in a network. It is a key ability in providing QoS, capacity planning and efficient traffic engineering. Furthermore, heavy flow detection is crucial for the detection of Distributed Denial of Service (DDoS) attacks in the network which remain a common attack in the Internet today, with hundreds of attacks carried out daily~\cite{Kaspersky}.

We present techniques for large flows detection in traffic that flows through an SDN switch
with Openflow. 
 While SDN switches are very efficient and considerably simpler to manage than existing routers and
switches, they do not offer direct means for the detection of large flows.

Existing network monitoring tools for classic IP networks have been available for over $20$ years, with one of the earliest tools being Cisco Netflow~\cite{Netflow}.
Over the years, traffic visibility, and specifically measurements and monitoring in IP networks has become an increasingly difficult task due to the overwhelming amounts of traffic and flows~\cite{ZhaoGWX06}.
While existing tools may be very useful for classic networks, monitoring in SDN networks requires new tools and technology. The SDN network architecture places the controller as the focal point of the network. Therefore, using existing tools would require extensive communication between the controller and the monitoring tools, which would place significant overhead on the controller. It is therefore necessary to provide new monitoring methods for SDN networks based on the SDN architecture.

We design ways to implement monitoring methods with the widespread OpenFlow and the recent P4 standard for SDN switches.
OpenFlow switches provide counters of the number of bytes and packets per flow entry, yet traffic measurement remains a difficult task in SDN for two reasons: First the hardware (usually Ternary Content Addressable Memories (TCAMs)) constraints limit the number of flows which the switch can maintain and follow. Secondly the limited number of
updates that the switch can process per second~\cite{StephensCFDC12}, which hence limits the number of
updates that the controller can make to the flow table.
The algorithms provided herein overcome these limitations by providing efficient building blocks for large flow detection and sampling which may be used by various monitoring applications.

\subsection{Our Contribution}

First, we propose our \algfull algorithm which is an efficient method to detect large or heavy flows going through an SDN switch.
The \algfull algorithm is designed for protocols which are based on the match and action model (e.g., OpenFlow, $P4$, etc.), 
and performs a division of labour between the switch and the controller, coordinating between them to identify the large flows. \algfull achieves very high accuracy using 
a fixed amount of rules in the switch and requiring little communication between the switch and the controller.

Second, as part of our algorithm we present various OpenFlow based methods to sample packets 
that traverse an SDN switch. These methods may be used independently of our heavy flows detection algorithm  

Third, we consider a distributed model with multiple switches and propose solutions for efficient scaling of our techniques, to support large flow detection as well as sampling in the distributed setting.

Finally we have implemented and evaluated our Sample\&Pick comparing it with OpenSketch~\cite{Opensketch}. The sampling methods rely on standard and optional features of OpenFlow 1.3 (or the $P4$ language) and are implemented with the NoviKit (hardware) switch\cite{NoviKit} (operated with NoviWare switching software \cite{NoviWare}). The heavy flow detection also relies on a standard OpenFlow controller and was evaluated as a whole using a dedicated virtual time simulation for both the data and control planes.
Additionally, the techniques presented are
both flow-table size and switch-controller communication efficient.

\section{Heavy Flows Detection}

\subsection{Background}\label{Section:FreqItems}

The problem of finding the heavy hitters or frequent items in a stream of data is as follows: given a parameter $v$ and a sequence of $N$ values $\alpha=\langle \alpha_1,.....\alpha_N \rangle$, using $O(v)$ space, find at most $v$ values each having a frequency (the number of times it appears in $\alpha$) which is greater than $\frac{N}{v}$.

Many solutions have been proposed for the heavy hitters problem, for example ~\cite{MetwallyAA05,MisraG82,AlonMS99,CormodeM04,
MankuM02}. A description of a few counter-based algorithms as well as other results regarding the heavy hitters problem can be found in~\cite{CormodeH08}.

We chose the algorithm of Metwally et  al.~\cite{MetwallyAA05} as a building block for the detection of heavy flows due to its simplicity, efficiency and high level of counter accuracy.
The (additive) error rate $\epsilon$ of this algorithm is $\epsilon = \frac{N}{v}$~\cite{MetwallyAA05}. Therefore, for each item $j$, denote its counter in the output of the algorithm $c_j$, and 
its count in the sequence as $r_j$ then $r_j \leq c_j \leq r_j + \epsilon$. 
The algorithm requires $O(v)$ space and 
only a single pass over the input, with a small number of instructions per item.

\subsection{Definitions}
Following \cite{OPenflowWhitePaper08} a \emph{flow} is defined to be any sequence of packets which can be matched to rules in the flow table, such as, for example, those defined by a set of header field values. Note that our algorithms can be used for any flow definition, including those which pertain to matches in the payload or any of the headers as long as it is supported by the controller and switch implementation.
A flow entry in OpenFlow (flow) table can be defined to match packets according to (almost) any selection of header field bits thereby allowing various flow definitions.

 A \emph{large flow} is usually defined as a flow that takes up more than a certain percentage of the link traffic during a given time interval~\cite{EstanVarghese03}.
For some applications other definitions of large flows are required,
for instance network analysis tools may want to identify flows that
consist of a certain amount of packets regardless of link
capacity. Therefore we refine the large flow definition, 
considering both the time
aspect as well as the type of measurement performed.

\Xomit{
 different types of large flows: }
We consider the following definitions of large flows:


\begin{definition}
\textbf{Heavy flow:} Given a stream of packets $S$, a \emph{heavy flow} is a flow which includes more than $T$ percent of the packets 
since the beginning of the measurement.
\end{definition}

Considering the definition of \emph{flow} provided above, this can be useful for identifying flows which remain heavy over a
significant period of time, for example in Distributed Denial of
Service (DDoS) attacks. On the other hand this will miss large
flows if the measurement continues for a very long period of time.

\begin{definition}
\textbf{Interval Heavy flow (Elephants):} Given a stream of packets $S$, and a length of time $m$, an \emph{interval heavy flow} is a flow that includes more than $T$ percent of the packets seen in
the previous $m$ time units.
\end{definition}
This can be used for standard traffic management and resource allocation.

\begin{definition}
\textbf{Bulky flow at a point of time:} Given a stream of packets $S$, and a length of time $m$, a \emph{bulky flow} is a flow that contains at least $B$ packets in the previous $m$ time units.
\end{definition}
\Xomit{
\begin{center}
\begin{table}[h]
\begin{tabular}{|l|p{2.4cm}|p{2.1cm}|}\hline
\backslashbox[25mm]{count type}{time}
& Limited Time
 Interval & Unlimited time \\ \hline
Percent of traffic & Interval Heavy flow & Heavy flow \\\hline
Amount of traffic & Bulky flow & ---- \\\hline
\end{tabular}
 \caption{Matrix definitions for large flows}
    \label{table:HeavyFlowDefinitions}
    \end{table}
\end{center}

}


The algorithms we present for large flows follow the above definitions which consider traffic volume measurements in terms of packets. Nevertheless, we note that certain traffic management capabilities require volume, i.e., byte size, analysis. For instance, if we wish to identify the flow which takes up the most bandwidth, then we are required to count the number of bytes in the flow rather than the number of packets.  The algorithms presented here work well for both definitions.

\subsection{Towards a Solution in SDN}
Fundamental counter based algorithms for finding Heavy Hitters (or flows) such as that of Metwally et. al.~\cite{MetwallyAA05}, cannot be directly implemented in the SDN framework since in the worst case they would require rule changes for every packet that traverses the switch. A different approach is therefore needed.

First we consider a naive solution which we name \alglightNoSp, that samples packets in the switch and then sends all sampled packets to the
controller. The controller computes the heavy flows using a heavy hitters algorithm. However, as can be seen in Figure \ref{fig:false-neg-errors} (and other works ~\cite{EstanVarghese03}), relying solely on the samples is not accurate enough. Next we consider a solution based on the \emph{\algsnh} paradigm of
~\cite{EstanVarghese03} which was devised for
identifying elephant flows in traffic of classic IP networks. In \algsnh sampled packets are sent to the controller, which installs a counter rule for each new flow that is sampled. Every consequent packet from that flow will be counted by the rule and will not be sampled.
By using sampling together with accurate in-band counters for sampled flows \algsnh achieves very accurate results, yet the high amount of counters and the rate of installing them make \algsnh incompatible with SDN switch architecture. Therefore we only consider it as a reference point to evaluate our algorithm.

\Xomit{
Our SDN based implementation of Sample and Hold, which we name \algsnhNoSp, 
uses the Devoflow~\cite{devoflow} infrastructure. It samples packets and then uses Devoflow to install macro rules which create a "hold" rule in the flow table for every new flow that is sampled, so that every subsequent packet in the flow can be counted. The problem with this solution is that it may install an unlimited number of rules in the switch.
}

To deal with the problems of the above solutions, we present our \algfull algorithm. \algfull uses sampling 
to identify flows that are suspicious of being heavy. For these
\emph{suspicious} flows a special rule is placed in the switch flow
table providing exact counters for them. The \algfull algorithm considers both the bounded rule space in the switch as well as the time it takes for the controller to install a rule in the switch. Therefore we use two separate thresholds, the first, $T$, for determining which flows are heavy and a second lower threshold, $t$,  for detecting potentially large flows. This lower threshold allows us to install rules in the switch early enough to get an accurate count of the large flows, yet we do not install rules for too many flows that will remain small. The \algfull algorithm is described in detail in Section~\ref{Section:SampleAndPick}.

Table~\ref{table:heavyflows} depicts the conceptual differences and the resource
consumption overhead of the \algfull algorithm, the SDN
\algsnh algorithm and the \alglight algorithm

\begin{center}
\begin{table*}[htp]
\begin{tabular}{|p{1.9cm}||p{2.9cm}|p{3.7cm}|p{3.4cm}|p{4.1cm}|}\hline
Technique & Switch memory usage & Controller functionality  & Controller to Switch traffic & Switch to controller traffic  \\ \hhline{|=||=|=|=|=|}
\algfull & Sampling rules + at most $\frac{1}{t}$ count rules & Heavy hitters computation + counter aggregation  & Every interval at most $\frac{1}{t}$ new count rules & Sample of all non-hold packets + counters each interval.\\\hline
Sample\&Hold (OpenFlow variant)&  Sampling rules + $unlimited$ count rules & Counter aggregation & Every new sample create message with a new count rule & Sample of all non-hold packets + final counters. \\\hline
\alglight & Sampling rules  & Heavy hitters computation & None & Sample of \emph{all} packets \\\hline
\end{tabular}
 \caption{Comparison of the heavy flow detection techniques presented in this paper. Denote $t$ the threshold for candidate heavy hitter in \algfull.}
    \label{table:heavyflows}
    \end{table*}
\end{center}



\Xomit{
Our \algfull algorithm, samples the flows in the network
to identify flows that are suspicious of being heavy. For these
\emph{suspicious} flows a special rule is placed in the switch flow
table so that we can provide exact counters for the suspicious
flows. Our method is based on the \emph{Sample and Hold} paradigm of
Estan and Varghese~\cite{EstanVarghese03} which was devised for
identifying elephant flows in the traffic. We also make use of the Heavy Hitters algorithm of Metwally et. al.~\cite{MetwallyAA05}.
}

\subsection{The \algfull algorithm}\label{Section:SampleAndPick}

\subsubsection{Algorithm Overview}

\begin{figure}[h]
\begin{center}
\includegraphics[trim = 0mm 40mm 0mm 21mm, clip, scale=0.30]{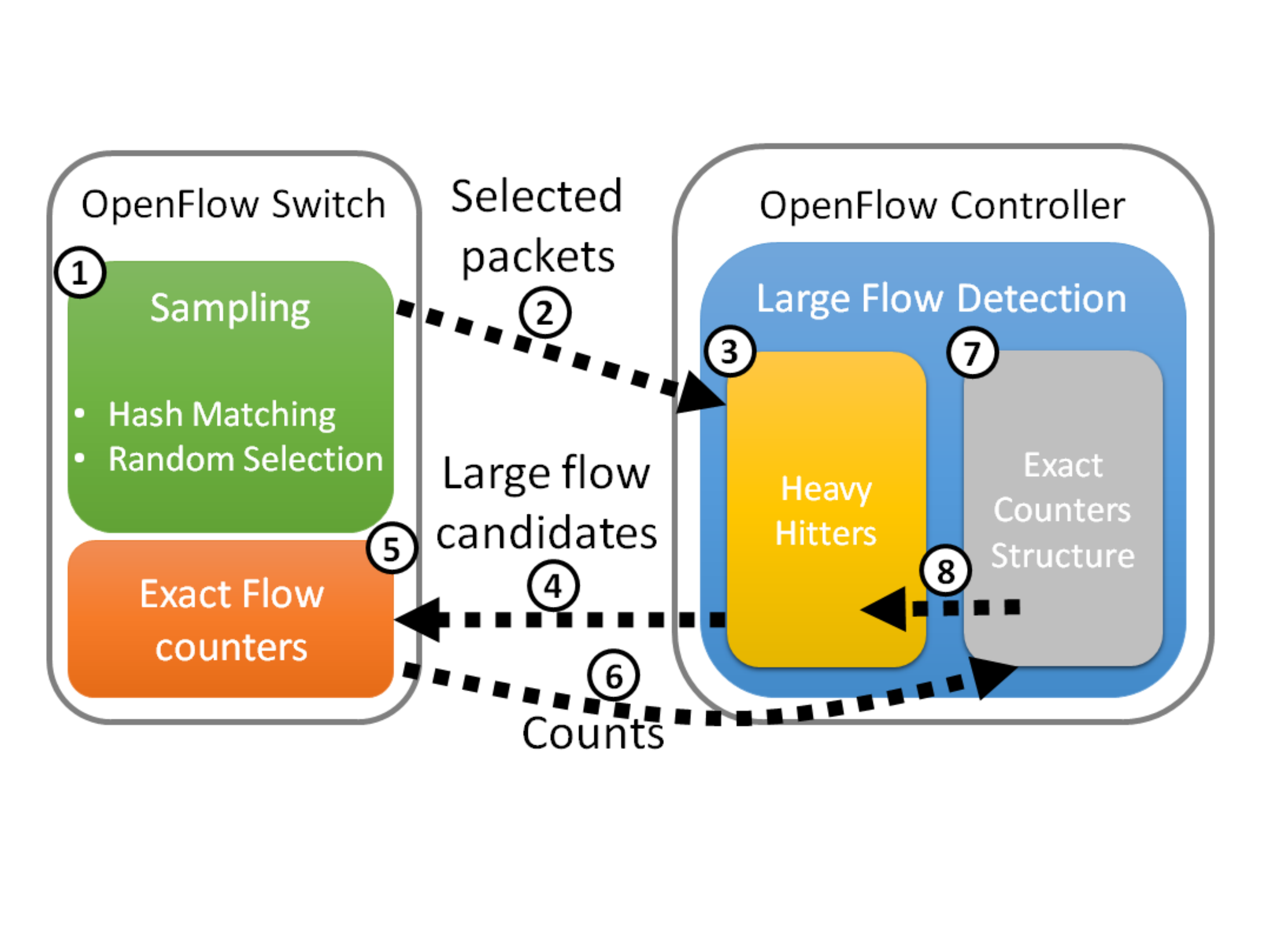}
\end{center}
\caption{\algfull overview}
\label{Figure:HeavyFlowOverview}
\end{figure}

Our algorithm operates as follows: 
in the first step we sample the
flows going through the switch using one of the sampling techniques to be explained in Section~\ref{Section:SamplingTechniques}.
As can be seen in Fig.~\ref{Figure:HeavyFlowOverview}, these samples
are sent to the controller,
that feeds them as input to a heavy hitters computation module in order to
identify the suspicious heavy flows (steps 2 and 3). Once a flow's counter in the
heavy hitters module has passed some predefined threshold $t$, a
rule is inserted in the switch to maintain an exact packet counter
for that flow (steps 4 and 5). This counter is polled by the
controller at fixed intervals and stored in the controller (steps 6 and 7).
Finally the last step increments the counters that are processed by the Heavy Hitters module to maintain correct counters of non-sampled flows.

\Xomit{
Recall, we denote the heavy flow threshold $T$ and the suspicious heavy flow threshold $t$.
Notice that $t$ is normally substantially smaller than the heavy flow threshold $T$, e.g.,
$t=\frac{T}{2}$,
so that a precise count of the heavy flows may be
achieved. }


\subsubsection{Switch Components Design}
As seen in Fig.~\ref{Figure:HeavyFlowOverview}, two kinds of rules
are used in the switch flow tables.  The \emph{sampling} rules,
which are created as needed by the selected sampling algorithm as
described in Section~\ref{Section:SamplingTechniques}. And the \emph{counter} rules used for precisely counting packets of
potentially heavy flows. An example of this configuration can be seen in Table~\ref{tab:downstream-rules}.

\begin{table}[t]
\centering
\small
\begin{tabular}{|p{1.1 cm}|l|l|}
\hline
name & match & actions\\
\hline \hline
Count $flow_1$ &$(src\_ip, src\_port, dst\_ip, dst\_port) = flow_1$
& 1
\\\hline
... & ... & ...
\\\hline
Count $flow_m$ &$(src\_ip, src\_port, dst\_ip, dst\_port) = flow_m$
& 1
\\\hline
Sample & $(src\_ip, src\_port, dst\_ip, dst\_port) = *$
& 2
\\\hline

\end{tabular}
\caption{Illustration of switch flow table configuration. 
Rule priority decreases from top to bottom. Actions: 1- increment counter; 2 - apply sampling technique (goto sampling tables / apply group)}
\label{tab:downstream-rules}
\end{table}

%

First, each packet is matched against \emph{counter} rules.
In case of a successful match,, the relevant counter
is increased.  Only if the packet does not match any \emph{counter} rule, it is
matched against the sampling rules, and if the packet is selected by the sampling rules it (or only the headers) is sent to the controller. 
Counters of the \emph{counter} rules are only sent to the controller when polled by the controller.

\subsubsection{Controller Components Design}
(Fig.~\ref{Figure:HeavyFlowOverview}) The controller maintains the
heavy hitters computation module and a collection of the exact
counters accumulation.

The \emph{heavy hitters computation module}: Maintains the heavy
hitters data structure according to the algorithm of Metwally et
al.~\cite{MetwallyAA05}, as described in
Section~\ref{Section:FreqItems}.


As the heavy hitters module only receives sampled data which is sent
to the controller from the switch, the traffic of the heavy flows
which are not sampled is not inserted at all into the heavy hitters
and therefore it may seem as though the flows are no longer heavy.
To simulate the sampling of these heavy flows, when the controller
polls the switch for the updated counters, it uses those counters to
update the heavy hitters module accordingly. That is, we simulate a
sampling of the heavy flows by updating the heavy hitters module
with the number of new packets that have been counted since the
previous polling, multiplied by the sampling ratio $p$.  As noted this
mechanism saves a substantial amount of sample traffic from the
switch to the controller.

The \emph{exact count data structure}: The accumulated
counters of the flows that are suspected to be heavy are maintained in a simple
ordered data structure.
Its use is to compute
the delta from the previous time the counters were polled.  This
delta is then fed (with a factor) into the heavy hitter module.

\Xomit{ We note that to simplify our controller mechanism, it is
possible to independently match all incoming packets against both
the sampling rules and the counting rules in the switch. Therefore
completely separating the exact counters from the sampling process
which is used for deciding which hold rules should be put in place.
In this manner, any sampled packet is sent to the controller, even
if there exists a hold rule for the flow for which the packet
belongs. The problem with this simplification is that each sampling
of a heavy flow is sent directly to the controller in addition to
being sent in a bulk using the counter. Therefore, it may cause a
substantial increase in the amount of traffic that the switch sends
to the controller. }

An additional counter is maintained in the controller to count the
total number of items inserted into the heavy hitters module, which
is necessary to calculate the rates from the individual counters
inside the heavy hitter module.  At any point the heavy flows may be
identified as the flows in the heavy hitters module that have passed
the threshold $T$, relative to the total counter.

\subsubsection{Analysis}
\label{Section:LFAnalysis}

Here we discuss how to choose the parameters, $t$ and $v$ of
\algfull algorithm for given problem parameters, the threshold $T$
for heavy flow and the sampling probability $p$.

By definition, if a total of $N$ packets have passed so far, each
heavy hitter flow contains at least $TN$ packets.
%
%
%
%
Our controller receives each packet with probability $p$.  The
number of samples is then on average (or exactly depending on the sampling
method) $n:=Np$.
The number of packets sampled out of $x$ original packets is a
random binomial variable with average $xp$ and variance $xp(1-p)$.
When $x$ is high this converges to normal distribution with similar
parameters.
For normal distribution, w.h.p the random variable is within distance of $3$ times the standard deviation from the average.
Therefore the number of packets sampled from $x$ packets is w.h.p greater than $xp-3\sqrt{xp(1-p)}$.

Our scheme uses a threshold $t<T$, in order to detect possible heavy
flows that might be missed due to sampling errors. For a heavy flow
(with at least $T\cdot N$ packets) w.h.p at least
$TNp-3\sqrt{TNp(1-p)}$ packets are sampled. We need to set $t$ to
ensure that the above expression is higher than $t\cdot n$.  Thus,
\begin{equation}\label{eq:t}
t<T-3\frac{\sqrt{T(1-p)}}{\sqrt{Np}}
\end{equation}
Since $t$ must be a positive number, we get the following constraint on the flow weight (ratio) our scheme is expected to detect: $T^2-9\frac{T(1-p)}{Np} > 0$ which is valid when
\begin{equation} \label{eq:T}
T>9\frac{1-p}{Np}
\end{equation}

For example, assuming a line rate of $6\cdot 10^5$ packets per
second and a controller throughput of only a few thousands messages
per second, we need a sampling rate of at most $1:100$, i.e.,
$p<10^{-2}$.  Assuming that the tested interval is at
least $10$ seconds long, more than six million packets
pass through the switch during the interval, i.e., $N>10^6$. From
Equation \ref{eq:T} we get that the threshold, $T$, can then be
roughly $10^{-3}$ or more.

\Xomit{
 Alternatively we can derive the minimal duration, $I$,
required for stabilization of sample based heavy flow
identification. We assume a steady line rate $L$, so we can express
$N$ in Equation \ref{eq:T} by $N=IL$ and we get the following
constraint:
\begin{equation}\label{eq:I}
I>9\frac{1-p}{LTp}
\end{equation}

}

Next we consider the fact that the flows that are monitored by exact
counters are updated in batches (when reading the switch flow entry
counters). To make sure that their counters in the
approximate HH structure are not evicted between updates, we set the number of entries, $v$, to be high enough considering the
threshold, $t$, for monitored flows.

Next we show that by choosing $v=2/t$ the number of samples that
would cause the eviction of one of the monitored flows, that is a
flow that is located at the top part of the approximate heavy
hitters structure, is very high.

Assume we have $k$ monitored flows, the sum of their counters is at
least $k\cdot n\cdot t$.  The number of other values in the table is
$v-k$, and their sum is at most $n - knt$.  In order for the minimal
monitored flow to be evicted, all lower values in the table should
exceed it, i.e., all smaller counts need to become higher than $nt$.
Their sum should thus be at least $(v-k)nt$, increasing by at least
$(v-k)\cdot nt - (n-knt) = vnt-n$.  Since the counts change by the
number of incoming samples,
if we set $v=\frac{2}{t}$ then the number of new samples
received between batch updates should be as large as the number of
all samples received so far ($n$) which is highly unlikely.

\subsection{Traffic Sampling}
\label{Section:SamplingTechniques}
An SDN controller sets flow entries in the switch, a flow entry can match one or many flows but generates one statistical record for all matching flows. A controller has to install a flow entry per each separately monitored flow in real time by sending all unmonitored flows to the controller which in turn would install a specific entry for each. Monitoring flows in real time in the controller is infeasible due to controller computation speed constraints.
\Xomit{
An SDN controller sets flow entries in the switch, a flow entry can match one or many flows but generates traffic statistics for all matching flows combined. Therefore in order to monitor each flow alone, a controller would have to install one flow entry per each flow, either in advance for all possible flows or in real time by sending all unmonitored flows to the controller which in turn would install a specific entry for each detected flow. Monitoring all possible flows in advance is infeasible due to switch memory constrains while processing all flows in real time in the controller is infeasible due to controller computation speed constraints.

SDN switches can be defined to monitor and count specific flows but it is impossible to define them to monitor each possible flow separately in advance.
Furthermore, it is infeasible for an SDN controller to process all traffic going through the switch and to define monitoring rules to each existing flow in real time.
}
Therefore in order to find large flows in SDN networks, sampling has to be used to reduce the set of monitored flows.

We discuss two types of traffic sampling: packet sampling and pseudo byte sampling, for which we provide the following definitions respectively:
\begin{definition}
\textbf{Packet sampling:}
Select each packet in a stream of packets with probability $p$, $0 \leq p \leq 1$.

\end{definition}
Note that the number of packets sampled from each flow times $1/p$ is an estimation of the real number of packets in the flow (during the sampling period).

\begin{definition}
\textbf{Pseudo byte sampling:}
Select each byte in a stream of traffic with probability $p$, $0 \leq p \leq 1$.


\end{definition}
Practically, this translates to a packet size based sampling, where given a stream of packets, a packet of size $s$ bytes is selected with probability $1-(1-p)^s$. For small enough $p$, this can be approximated by $1-e^{-ps}$ and since usually $ps<<1$ it is approximated by simply $p\cdot s$.
With this type of sampling, the number of packets sampled from each flow times $1/p$ is an estimation for the real number of {\bf bytes} in each flow (during the sampling period).



\subsubsection{Packet Sampling}

\Xomit{
In \emph{Packet Sampling} we select each packet in a stream of packets traversing the switch with probability $p$, $0 \leq p \leq 1$, and send them to a \emph{receiver} that can be the controller or some middlebox (monitoring box).
}
We present two approaches for packet sampling, each using different SDN features.

\emph{Packet Sampling Using Random Selection:}
The following technique in the most direct way to implement packet sampling, it utilizes OpenFlow weighted groups (Section 7.3.4.2 in~\cite{OpenFlow132}), an optional feature intended for unequal load sharing and we expect it to be supported by future P4 compilers.

\Xomit{
The technique presented here is based on OpenFlow unequal load sharing with select groups \Xomit{weighted selection groups} (Section 7.3.4.2 in~\cite{OpenFlow132}).
}

A weighted group contains a list of buckets each with different weight and actions. A packet is assigned to such a group (by  the apply\_group instruction) is randomly diverted to one of the buckets
according to the weights and that bucket's actions are applied to the packet.

In our case, we use a group with two buckets -  an "active" bucket that transfers to the receiver and a "dummy" bucket does nothing. We set the weights of the buckets according to the sampling probability $p$:  weight $1$ for the active bucket and weight $\lceil\frac{1}{p}\rceil-1$ for the dummy bucket.

Note that as weighted groups are optional in the OpenFlow standard and are currently considered expensive (in terms of switch resources) and are not supported by P4, this solution is not compatible with all switches.

A similar sampling technique can be achieved by using OpenFlow round-robin groups, where  for each packet the next action bucket is chosen (in round robin order). This technique is less compatible and more expensive than wighted groups based technique, and we therefore only use it for comparison with other techniques without describing the full implementation details.

\emph{Packet Sampling Using Hash Matching:}
As random generators are not natively supported by current SDN standards, OpenFlow and P4, we suggest to use the hash of the packets instead. More precisely we suggest to use Ethernet CRC or TCP/UDP checksum fields and match them against predefined bit patterns thereby selecting which packets to sample and send to the collector. We overcome weaknesses of this method in the sequel.

More precisely, assuming $p=\frac{1}{2^k}$ the controller randomly selects a ternary pattern with $k$ $0/1$-bits (not '*'s) for matching the checksum field, and install a flow entry with that pattern as match and with an action to forward to the collector.
For example, sampling with probability $p=2^{-13}~0.0001$ is implemented by matching the ($16$bit) checksum to a ternary pattern with $3$ $*'s$ (don't cares) and $13$ zero/one bits.

Matching unconventional packet fields (e.g. checksum) is supported in P4
and is also supported by some SDN switches such as the NoviKit \cite{NoviKit,NoviWare} using the optional Experimenter extension.
In general this method uses the fundamental properties of all match-action modules (flow tables, TCAMs, etc..) and therefore expected to be easily realized in future network devices and control protocols.

Note that setting a single match pattern without changing it may present some problems. For example, crafted packets such as those
in DDoS attacks may be missed. Such packets may be generated with a specific checksum value, and would be missed by this method.
In order to deal with such scenarios, the controller should
modify the selected match pattern randomly every fixed period of time, so that the mechanism approximates a sampling with a uniform probability for selecting any packet over a long enough period of time.

Note that since each change in the bit pattern requires a new rule (e.g., OpenFlow FlowMod command) to be sent by the controller to the switch, there is a tradeoff between the safety of the scheme and
the control traffic it creates. It is also possible to send multiple commands in batches utilizing rule timeouts to set the end time of rule liveness, yet these rules have to be separated by additional rules to set the start time.
Considering a short $1$sec update interval and a command packet size of $108$ bytes ($40$ bytes for TCP/IP headers and $68$ bytes for OpenFlow 1.3 FlowMod message with two actions) we get an insignificant control plane traffic of $108$B/s (in each direction).

\begin{figure}[h]
\begin{center}
\includegraphics[trim = 0mm 0mm 0mm 0mm, clip, scale=0.2]{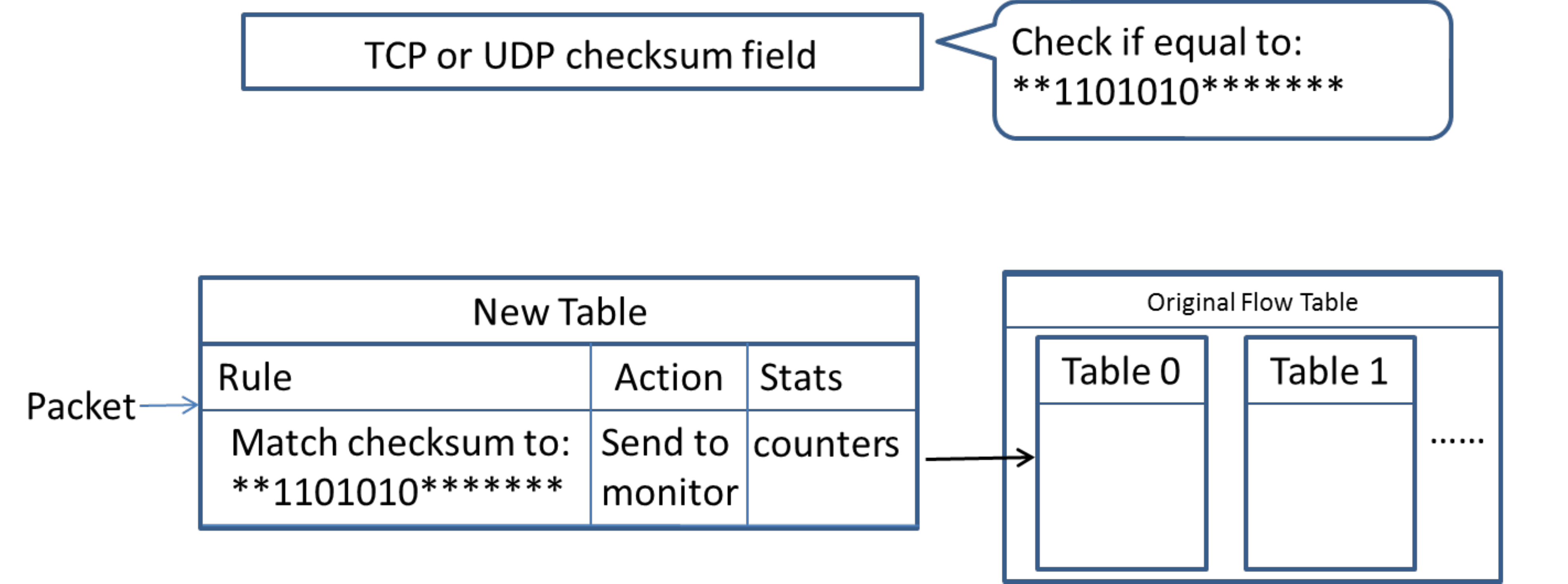}
\end{center}
\caption{Example of the \emph{randomized bit} algorithm for packet sampling. All packets traverse through both the new table and the original flow table. The sampling rate provided is $p=\frac{1}{128}$. Sampled packets may be sent to a monitor or the controller.}
\label{Figure:randombitSampling}
\end{figure}

Note that the flow entries can be installed in a dedicated flow table, so that the sampling process does not interfere with other switch processing. Packets that match the pattern are sampled and then \emph{all} packets continue to traverse the rest of the tables as in the unmodified pipeline. This process is depicted in Fig.~\ref{Figure:randombitSampling}.

\Xomit{
\emph{OpenFlow based Implementation:}
Assuming $p=\frac{1}{2^k}$ the controller randomly selects a ternary pattern with $k$ $0/1$-bits (not '*'s) for matching the checksum field.  The controller then creates a rule for the selected pattern and sends the rule to the switch. This new rule creates a new entry in a dedicated flow table, so that the sampling process does not interfere with other switch processing. Packets that match the pattern are sampled and then \emph{all} packets continue to traverse the rest of the tables as in the unmodified pipeline. This process is depicted in Fig.~\ref{Figure:randombitSampling}.
Every fixed time interval, the controller randomly selects a new bit pattern and sends the updated rule to the switch.}

\subsubsection{Pseudo Byte Sampling}
As described above, \emph{Pseudo Byte Sampling}  with probability $p$ per byte is approximated by sampling each packet with probability $p\cdot s$, where $s$ is the packet size.

We present optimized techniques for pseudo-byte sampling, which are based on matching the packet size.
Matching unconventional packet fields (e.g. packet size) is supported in P4
and is also supported by some SDN switches such as the NoviKit \cite{NoviKit,NoviWare} using the OpenFlow optional Experimenter extension

\Xomit{
we assume that packet size is accessible and can be matched in the OpenFlow pipeline, using the Experimenter extension or any other way.
It is expected that with upcoming new OpenFlow and alike SDN architectures, such as \cite{p4} more header fields could be natively matched by the flow tables.
}

\emph{A General Approach for Pseudo-Byte Sampling:}
A straightforward implementation of the pseudo-byte sampling is to use multiple instances of any of the packet sampling implementations presented so far, where each instance samples with a different probability, and we divert each packet to the most accurate sampling instance considering the packet size. More formally, given a set of packet sizes $\{s_i\}_{1\leq i\leq R}$, we define the set of sampling instances $\{PS_i\}_{1\leq i\leq R}$,
where $PS_i$ samples any packet with probability $p\cdot s_i$. Moreover, we divert each packet with size $s$ to the sampler $PS_z$, where $z=arg min_i {|s-s_i|}$.

The maximum error ratio in this method is $max_i \frac{s_i}{s_{i-1}}$. Therefore, in order to bound the error $s_i$s should be chosen as geometric series. For example, for $1\leq i\leq R$,  $s_i=m\cdot 2^i$ where $m$ and $M$ are min and max packet size (e.g., 64 and 1500 for Ethernet) and $R=\log_2\frac{M}{m}$.
Finally, following last example, given a packet of size $s$ we divert it to the $PS_{\lceil\log_2 s\rceil}$.

\Xomit{
However this approach well suits only the random selection based sampling where the resources per sampler instance are very low (only one group and two buckets per instance) but for the other methods this approach yields high resource complexity.
}

Note that this approach presents a tradeoff between accuracy and resources. In order to reduce the maximum error, one has to use more sampler instances.

\emph{Pseudo-Byte Sampling with Hash Comparison:}
The following sampling technique uses constant resources and has optimal accuracy. It is fully supported by P4  and is also supported by some SDN switches such as the NoviKit \cite{NoviKit,NoviWare} using the OpenFlow optional Experimenter extension.

Before describing the technique we first make the following observation: if $s,M$ are numbers such that $0<s<M$ and $x$ is a random variable chosen uniformly from $[0,M]$, then the probability that $x\leq s$ is $s/M$, i.e. for $x\sim U([0,M]), Pr(x\leq s)=\frac{s}{M}$.
Following last observation, if we substitute $M$ with $\frac{1}{p}$, we get that the probability of sampling a packet of size $s$, namely $ps$, is equal to the probability that $x<s$. This means that given access to such uniform distribution we can implement size based sampling in the following way: for each packet of size $s$, first randomly choose $x$, then if $x<s$ transfer the packet to the receiver.

Similarly to the hash matching technique, we suggest to use the packet checksum as a random number generator.
\Xomit{
However, since OpenFlow doesn't include random number generator, our technique uses the checksum field as such.
}
Assuming $\frac{1}{p}=2^b$, where $b\in \mathbb{N}$, we use the first $b$ bits of the checksum field as the random variable $x$, and we define rules that checks whether $x<s$. If the comparison succeeds the action should forward the packets to the receiver and otherwise do nothing.

\Xomit{
First we observe that sampling a packet of size $s$ with probability $ps$ is equal to sampling this packet if a random variable $x$ uniformly distributed in $[0,\frac{1}{p}]$ is smaller than $s$, i.e. for $x\sim U([0,\frac{1}{p}]), Pr(x<s)=ps$.
Secondly, assuming $\frac{1}{p}=2^r$, where $r\in \mathbb{N}$, we consider the first $r$ bits of the checksum field as the random variable $x$, and we define rules that checks whether $x<s$. If the comparison succeeds the action should forward the packets to the controller and otherwise do nothing.
}
Comparing two fields is also not natively supported in OpenFlow but can be implemented by a flow table filled with $2b+1$ rules, where $b$ is the width of the compared numbers (in bits) \cite{liron-ranges}.

Similarly to the packet sampling with hash matching described in the previous section,
the pseudo-byte sampling technique presented here might miss specific classes of crafted packets whose checksum is high. Therefore we need to add some external randomness that is changed over time. While in the case of packet sampling
we can just change the pattern, in the pseudo byte sampling case we need to affect the comparison result.

The solution we suggest is that every fixed time interval the controller will modify a rule (or a batch of rules with different timeouts) that writes the metadata field of every packet with some value $r$, and that value will be used in a modified version of comparison that checks whether $x\oplus r<s$. For each new value of $r$ the controller needs to send one FlowMod command packet whose size in our scenario is less than $110$ bytes. As in the packet sampling case, even for short time interval of $10$ seconds, the control traffic overhead is insignificant.

\subsection{Evaluation}
\label{Section:Evaluation}
\Xomit{
\emph{Sampling:}
We  evaluate the performance of our sampling schemes by considering
the resulting sampled flow size distributions
compared to real flow size distribution. Figure \ref{fig:cdf} shows the three sampling methods achieve
similar distributions, and closely approximate the real (exact)
distribution.
}



We compare our \algfull algorithm to the two additional solutions described above \algsnh and \alglightNoSp (See algorithms overview in Table~\ref{table:heavyflows}. 
We analyze the resource consumption and accuracy of each of the algorithms in fixed time intervals. We use $10$ intervals of $5$ seconds each, and we collect the counters of each algorithm at the end of each interval.
In addition we compare the results of these algorithms to that of the OpenSketch Heavy Hitters detection mechanism~\cite{Opensketch}.
For our analysis, we use a one-hour packet trace collected at a backbone link of a Tier-1 ISP in San Jose, CA, at 12pm on September 17, 2009~\cite{caida2009}.

We chose the following simulation parameters $T=5\cdot10^{-3}$, $p=\frac{1}{1024 \cdot 10^{2}}$ Bytes, $t=2\cdot10^{-3}$, $v=2000$. 
\Xomit{
\begin{figure}[h]
                \includegraphics[clip=true,trim=0mm 0mm 0mm 10mm, scale=0.4]{FalseNegatives_caida12_shir.pdf}
                \caption{False negative errors, shown by the ratio between the Heavy Hitter (HH) flows missed to total number of HH flows.}
                \label{fig:false-neg-errors}
\end{figure}

In the above example, it takes the scheme about $3\cdot10^7$
packets to stabilize (Figure \ref{fig:false-neg-errors}).

\begin{figure}[h]
                \includegraphics[clip=true,trim=0mm 39mm 0mm 0mm, scale=0.33]{Shir-comm-in+7-2Z.pdf}
                \caption{Rate of PacketIn messages (samples) from switch to controller. In \algsnhNoSp, sampling is switch-contained.}
  \label{fig:comm-in}
\end{figure}
}

\begin{figure*}
       \centering
       \begin{subfigure}[b]{0.49\textwidth}
                \includegraphics[clip=true,trim=0mm 0mm 0mm 0mm, scale=0.16]{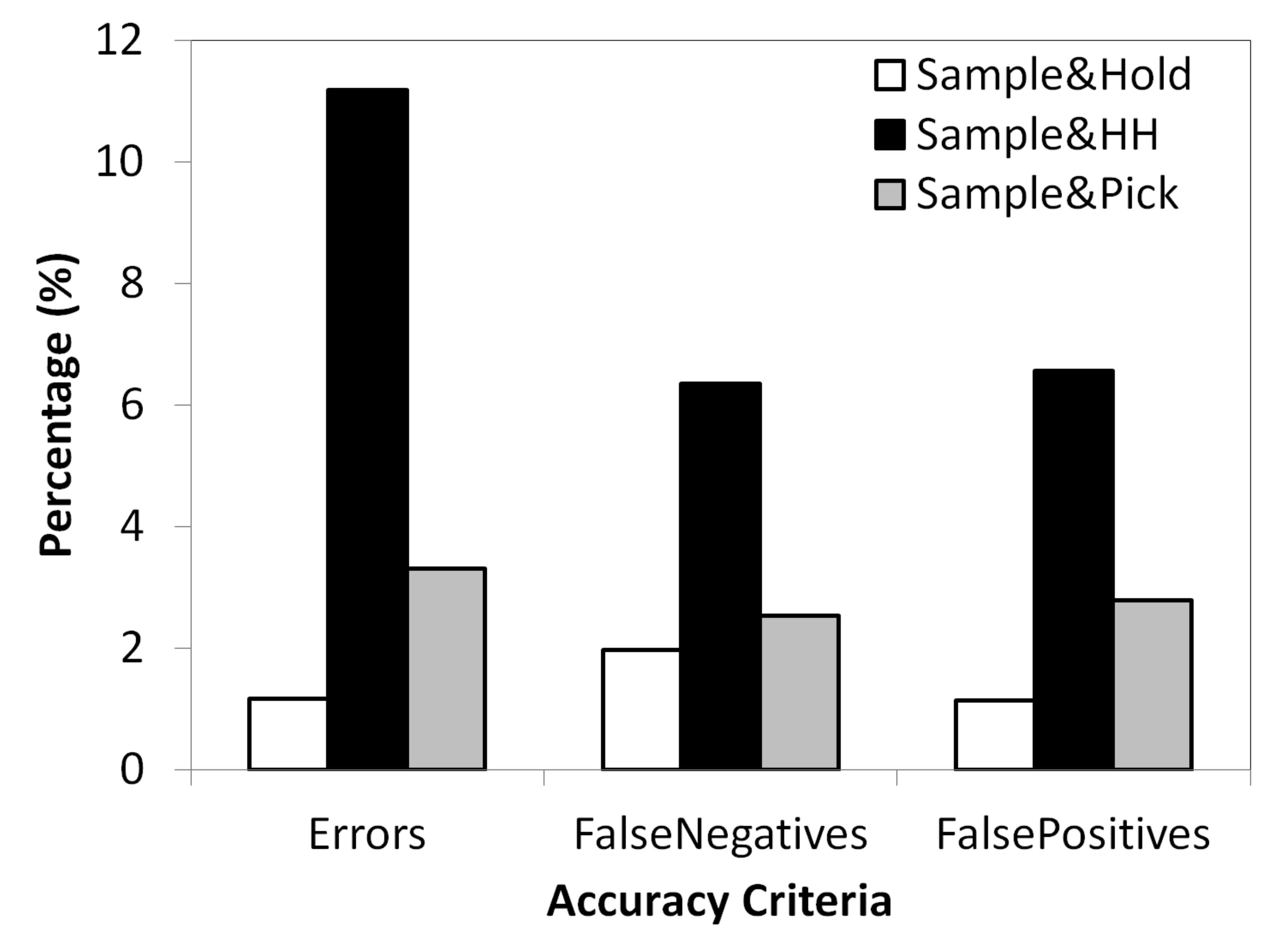}
                \caption{Comparison of algorithms by Counter error, False negative errors and False positive errors. }
                \label{fig:false-neg-errors}
        \end{subfigure}
        \hfill
        \begin{subfigure}[b]{0.49\textwidth}
                \includegraphics[clip=true,trim=0mm 0mm 0mm 0mm, scale=0.16]{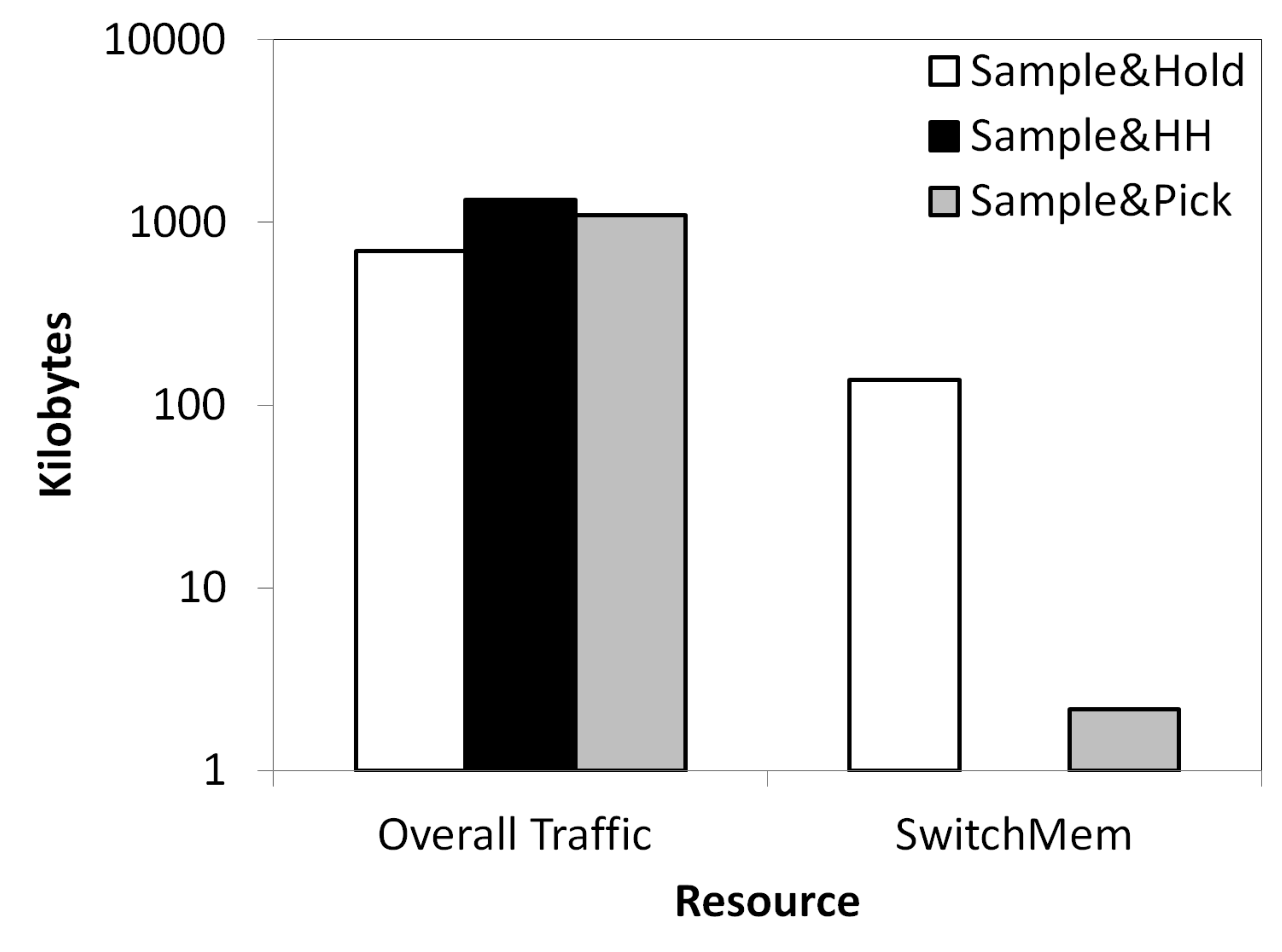}
                \caption{Comparison of algorithms by Overall traffic (between switch and controller) and Switch memory usage.}
  \label{fig:comm-in}
\end{subfigure}
        \caption{Resource consumption and accuracy comparison}
\end{figure*}

Figure~\ref{fig:false-neg-errors} shows a comparison of the three algorithms based on accuracy criteria. The counter error refers to the ratio between the real count of the heavy hitters and the algorithm estimates. The false negative and false positive errors is the ratio between Heavy Hitter (HH) flows missed to the total number of HH flows, and the HH flows wrongly detected to the total number of HH flows respectively. Figure~\ref{fig:comm-in} shows a comparison of the three algorithms based on the amount of traffic they generate and the amount of memory they use in the switch. As can be seen, while \algsnh provides the best accuracy results, it requires an increasing amount of counters and therefore its switch memory consumption is significantly higher than that of the other algorithms. In contrast, \alglight requires the least amount of switch memory since all of the heavy hitters computation is performed in the controller yet it relies on sampling alone and provides significantly lower accuracy results. Our testing shows that the \algfull provides accuracy results only slightly inferior to those of \algsnh yet requires significantly less switch memory.

\begin{center}
\begin{table}[htp]
\begin{tabular}{|p{2.0cm}||p{1.5cm}|p{0.8cm}|p{1.0cm}|p{1.5cm}|}\hline
Technique & OpenFlow Compatibility & Error Rate & Switch memory usage & Controller $\leftrightarrow$ Switch Traffic\\ \hhline{|=||=|=|=|=|}
\algfull& Yes &  $3.3\%$ & $2KB$  & $220KB/s$\\\hline
Sample\&Hold& Yes & $1.15\%$  & $400KB$ & $140KB/s$\\\hline
\alglight& Yes &  $11.3\%$ & $\leq 1KB$  & $270KB/s$ \\\hline
OpenSketch ~\cite{Opensketch}& No & $0.05-10\%$ & $94KB-600KB$ & NA \\\hline
\end{tabular}
 \caption{Resource consumption test results}
    \label{table:evalSummary}
    \end{table}
\end{center}

As can be seen in Table~\ref{table:evalSummary}, \algsnh gives the smallest error rate, since it performs an actual count of all flows that it samples, yet it uses significantly more switch memory. \alglight uses only samples for the counter estimates without using any counters in the switch yet incurs significantly higher error rates. \algfull has relatively small error rates due to the actual counting of potentially heavy flows, yet due to the careful selection of which counters to place in the switch, the switch memory usage in \algfull is very low. According to our testing, the error rate of \algfull may be further reduced with increased sampling rate or counter polling rate, yet the switch memory requirement remains steady at $2KB$ as determined by our parameters.
The controller$\leftrightarrow$switch traffic (sum of traffic in both directions) of each of the presented algorithms is directly influenced by the sampling rate (recall in this case $p=\frac{1}{1024 \cdot 10^{2}}$ Bytes) and the counter polling rate of the controller. In the case of \algfull the polling rate is set to be every $0.1$ seconds in these tests, while in \algsnh the controller only polls for the counters once at the end of the interval. As can be seen, \alglight produces a larger traffic overhead since all sampled messages are sent to the controller whereas in the other two algorithms the counters in the switch perform the aggregation locally.

Additionally, we compare our results to testing done on the OpenSketch Heavy Hitters detection mechanism~\cite{Opensketch}. OpenSketch is a very efficient measurement architecture, yet it is not compliant with the OpenFlow standard. Our \algfull algorithm was designed with the current OpenFlow and P4 ablities in mind and it can therefore be implemented using the current standards. We base our comparison on the evaluation results shown in~\cite{Opensketch}. Note that while we perform our test on the same data as used in~\cite{Opensketch}, we provide an average of $10$ intervals of $5$ seconds each, as opposed to $120$ intervals used in the OpenSketch evaluation. As can be seen in Table~\ref{table:evalSummary}, 
\algfull requires very little switch memory while achieving counter errors which are similar to those achieved by OpenSketch which uses significantly more switch memory.  The traffic overhead for OpenSketch is not provided in~\cite{Opensketch} and therefore we do not indicate it.

\section{Interval Heavy Flow and Bulky Flow Detection}

Recall that, an \emph{interval heavy flow} is a flow whose volume is
more than $T$ percent of the traffic seen in the last time interval
of length $m$.  While the problem is defined in a continuous manner,
that is, an interval can begin at any point in time, considering the
inherent subtle delays caused by the OpenFlow architecture, an
approximate solution is sufficient.

\begin{figure}[h]
\begin{center}
\includegraphics[trim = 5mm 110mm 100mm 25mm, clip, scale=0.40]{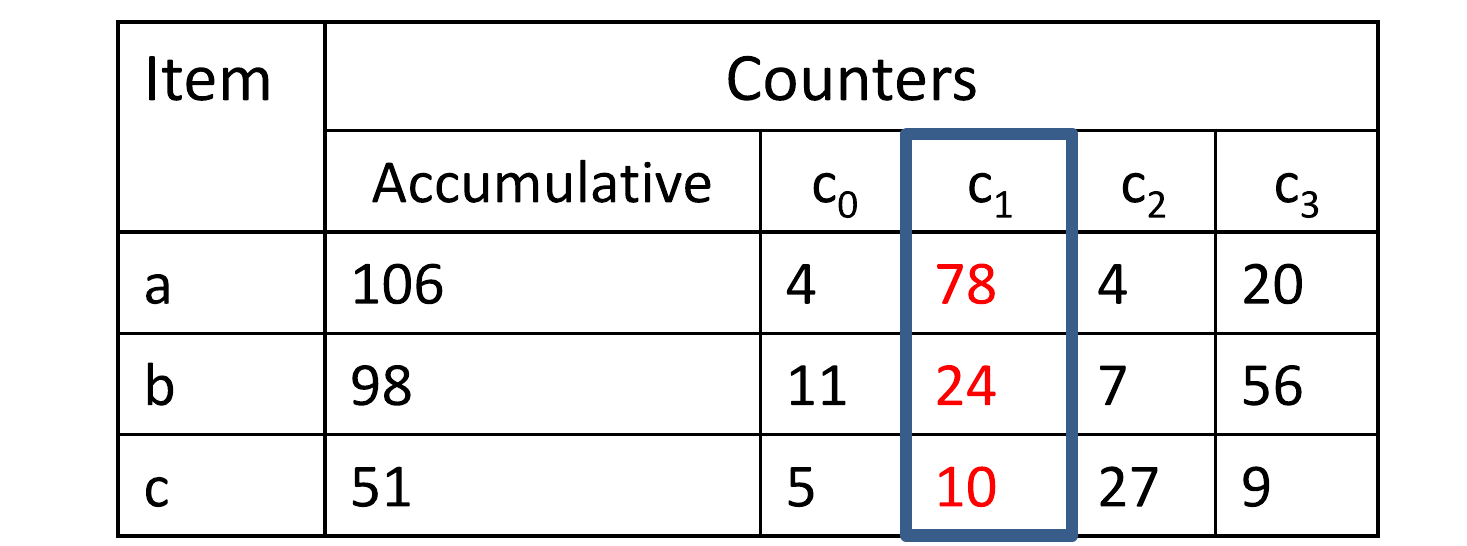}
\end{center}
\caption{The modified heavy hitters data structure using counter arrays. In this example the active counter is currently $c_1$.}
\label{Figure:intervalHF}
\end{figure}

Our solution 
makes use of the \algfull algorithm,
specifically we take the array of counters in the heavy hitter
module in the controller as the starting point. We modify this structure so that instead of
maintaining one counter per item (flow), an array of
counters is maintained for \emph{each flow} that is kept in the heavy hitter module. In addition, for each flow we maintain an additional accumulative counter. The updated counter structure is depicted in Fig.~\ref{Figure:intervalHF}.

The array of counters for each flow maintains the history of the flow's counter values in fixed intervals of time. The flow's accumulative counter is the sum of all the counters in the flow's array.  Let $m$ seconds be the
selected time interval, and let there be 
$r$ history counters maintained for each flow, we get a sub-interval that is
$\frac{m}{r}$ seconds long.  The basic idea is that in each
sub-interval a different counter in the array is updated by the HH
module, in addition to updating the accumulative counter. Thereby, consecutive (cyclicly) counters in the array can be
used to calculate the number of times the value appeared in the
entire interval.  At the beginning of the sub-interval, for each flow, the value of the active counter is decreased from the value of the accumulative counter. Then all active counters in all flows are reset to zero.  In this manner, at the end of each sub-interval, for any flow, the active counter equals the number of times the flow was sampled during
that sub-interval, and the value of the accumulative counter equals the number of times the flow was sampled in the last interval $m$. It follows that if the index of the active
counter is $a$ s.t. $0 \leq a \leq r-1$ for any $r' \leq r-1$ the sum of the cyclically consecutive counters between index $a - r' \mod r$ and $a$ equals to the number of times the item was seen during the $r'$ previous sub-intervals.

Note that if an interval does not begin at the beginning of an exact sub-interval, we will consider it to begin at the start of either the current or the consequent sub-interval.

The accumulative counter has two additional important uses: 1) it is used to maintain the threshold ratio; 2) it is used by the heavy hitters algorithm as the de-facto counter for deciding which flow has the minimum counter and should be evicted.

Using the accumulative counter in this manner is the basis for the correctness of our algorithm, which we will now briefly show. Given an interval $i$ of length $m$, denote $N$ to be the number of items seen in $i$. 
If $i$ is made up only of whole sub-intervals, it is easy to see that at the end of interval $i$ the accumulative counter of each flow in the structure is equal to what its counter would be had we reset all of the counters at the beginning of the interval. Therefore, using the accumulative counters as described above provides us with a heavy hitters mechanism which supports the same counter error rate (i.e. $\frac{N}{v}$) as that of~\cite{MetwallyAA05}. If, however, $i$ begins in the middle of a sub-interval, the counter error rate is slightly higher. In this case, suppose $i$ contains $j$ complete sub-intervals, and at most $2$ partial sub-intervals. The additional error contains appearances of the flow which occurred in the partial sub-intervals, which may incur an additional error of at most $\frac{N}{v}$ since otherwise it would be heavy for an interval comprised of only complete sub-intervals as well, making the overall error rate in this case to be $\frac{2N}{v}$.

Notice that bulky flows can be detected by using the above mechanism
without dividing the counters sum by the relevant sum of counters, but rather taking the absolute values.

\section{Distributed Setting}\label{Section:distributed}

In many cases, in order to achieve a comprehensive view of the network, it is required to distributively monitor traffic at multiple switches.
There are two main challenges to deal with when detecting large flows in this distributed setting; false negatives due to split flows and false positives due to sequential flows. Split flows are large flows that their traffic is split to small sub flows, each going through a different monitoring switch, and therefore monitored in parallel. Sequential flows are small flows that each of their packets traverse multiple monitoring switches and are therefore over sampled or counted.

In this section we extend our \algfull solution in order to support this distributed setting. We describe the changes that need to be done to the sampling and to the large flow detection scheme. We note that our solution easily scales with the number of monitoring switches. To support multiple controllers, a hierarchy of controllers needs to be defined and data should be collected by the controllers and forwarded up the hierarchy.

\emph{Sampling:}
In order to handle over sampling of sequential flows, flows that each of their packets traverse multiple switches, we need to prevent each packet from being sampled more than once. We suggest to do so by marking packets after they are sampled (whether selected or not) and by applying sampling only to unmarked packets.
Marking of packets can easily be managed in SDNs (with OpenFlow and especially with P4), for example by utilizing one bit in the VLAN tag. Matching the VLAN tag of each packet can be easily done and allows to skip sampled packets.
Note that the marks should be removed at egress ports so that they do not affect the traffic leaving the network.

\emph{Heavy Flow Detection:}
As described in Section~\ref{Section:SampleAndPick}, our \algfull algorithm makes use of both sampling and exact counter rules in the switch. To support the distributed setting, and to handle
split flows, that each of their sub flows goes through a different monitoring switch, all of the samples and counter values from all monitoring switches should be aggregated centrally by the controller. The controller will receive the samples and counter values from the different switches and treat them as if they were generated by a single monitoring switch. One of the implications of that is that when a flow becomes suspect of being large, exact counter rules should be installed on all monitoring switches, to assure that all consequent packets going through the network are counted.

\begin{figure}[h]
\begin{center}
\includegraphics[clip = true, trim = 0mm 0mm 0mm 0mm, width=\linewidth]{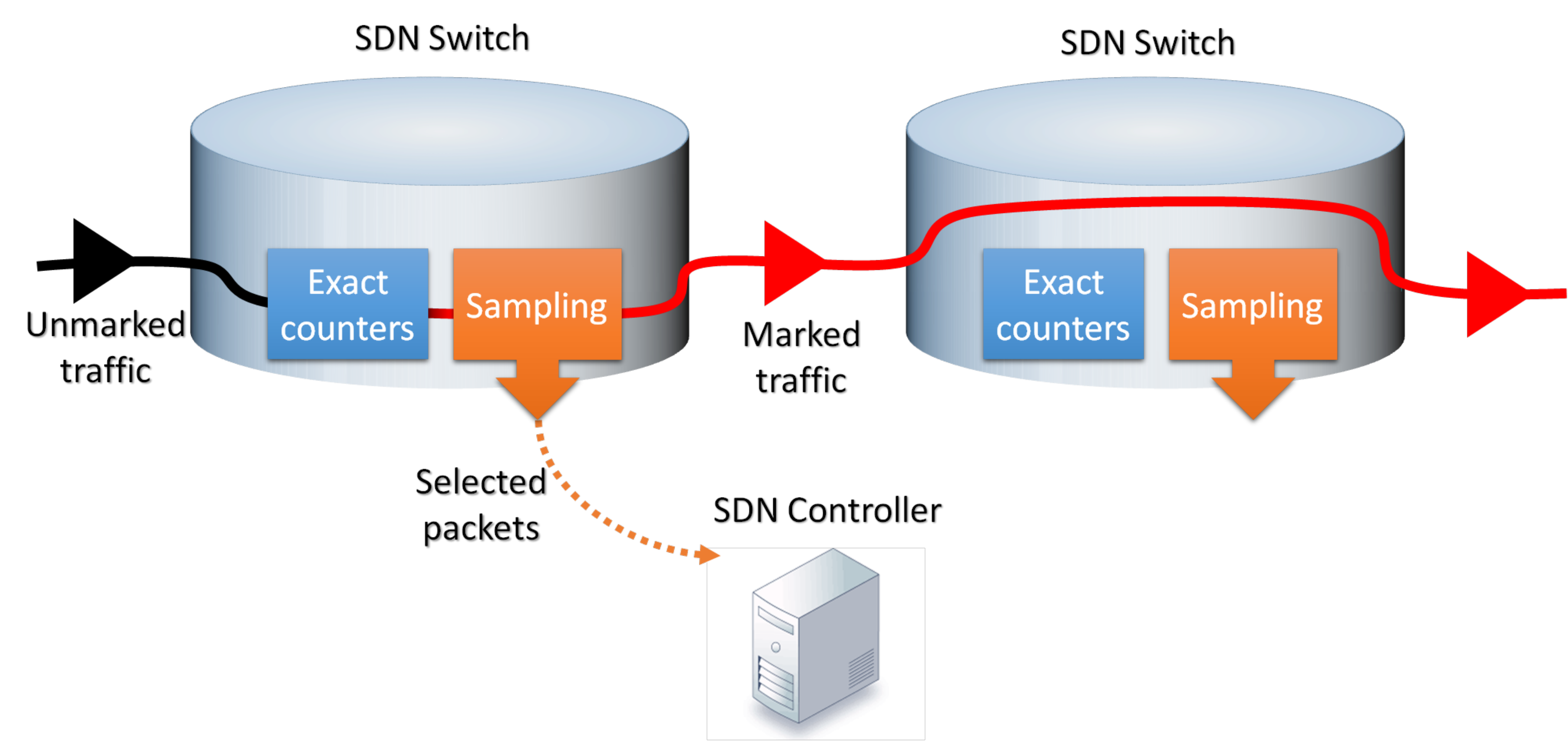}
\end{center}
\caption{Marking sampled packets in the distributed setting.}
\label{fig:marking}
\end{figure}

Similarly to sampling, in case of sequential flows that traverse multiple switches, exact counters (on different switches) should not count the same packet more than once.
The same packet marking technique we suggest to avoid over sampling, can be used in order to prevent multiple counting (see Figure \ref{fig:marking}), i.e., marked packets are not matched against exact counter rules nor sampled. Moreover, packets which match exact count rules are marked even if they have not been sampled.

\section{Related Work}
\label{Section:Related}
One of the earliest network monitoring tools was Cisco Netflow~\cite{Netflow}, which allowed collection of IP flow level statistics. Netflow provided 
the ability to gather information from the router about every IP flow, including byte and packet counts yet suffered from high processing and collection overheads, which were partially decreased using sampling in the variant \emph{Sampled} Netflow, yet this variant provided reduced accuracy caused by the straightforward use of sampling~\cite{EstanVarghese03}.
In~\cite{EstanVarghese03} Estan and Varghese significantly improve the accuracy of the sampling process by introducing the \emph{Sample and Hold} algorithm which provides better accuracy while reducing the processing and collection overhead. The sample and hold algorithm is essentially sampling with a "twist". As in regular sampling, each packet is sampled with some probability, and if there is no entry for the packet's flow, an entry is created. Once an entry for a flow exists, it is updated for every packet thereafter in that flow.

In a usual setup, monitoring devices are placed in central locations in
the network (such as Arbor's Peekflow \cite{peekflow}, or other
security detection devices) and samples of traffic are being sent to
the monitoring devices for various additional processing for which the switch/router are not suitable, such as heavy-hitters analysis, DPI, and
behavioral analysis. These monitoring devices usually cannot absorb and
process all the traffic. Therefore, traffic must be sampled, and only the
samples or relevant flows should be forwarded to these devices.


As the networks evolved, network monitoring tools with more advanced capabilities were developed. In~\cite{csamp}, for example, a flow monitoring tool was presented. There, they discussed adding flow sampling abilities as an inherent capability of the routers. They provide a framework for distributing the monitoring across routers, allowing for network-wide monitoring. By using uniform hash functions, flow sampling is not duplicated across different routers which route the same flow.

In OpenFlow the flow table allows us to define rules which support counting of bytes and packets per flow. However, this is not sufficient for more advanced measurements.
Recently there have been several works that discuss or suggest enhancements to network measurement capabilities for both OpenFlow and for SDN in general. 
FleXam is a sampling infrastructure for OpenFlow proposed in~\cite{Shirali-ShahrezaG13}, which adds sampling capabilities, using random number generation.
Opensketch~\cite{Opensketch} provides a simple approach to collect and use measurement data, separating the measurement data plane from the control plane.  The paper suggests a new architecture, where in the data plane, a pipeline of three essential building blocks is provided: hashing, filtering and counting, and in the control plane, a wide library of measurement tasks is provided.
The above works suggest an alternate to the OpenFlow architecture, while our work relies on features
that already appear in the current OpenFlow standard as required or optional features, in addition to the common extensions of as matching on an extra field in the packet. These extensions follow the concepts described in~\cite{devoflow}, that suggests that the OpenFlow standard should allow the user to configure the headers that the switch can examine.
All our modification are in the spirit of OpenFlow architecture.

We note that there are works that do not require changes to the OpenFlow standard. For instance, OpenNetMon described in~\cite{OpenNetMon14} is a controller module for monitoring flow level metrics, such as packet loss, delays and throughput in OpenFlow networks, for determining whether QoS criteria are met, which is based on the OpenFlow standard. Our solution which combines both a switch module and a controller module provides accurate results while significantly reducing the communication overhead.

A recent work~\cite{DistMonSDN}, proposes a method for distributing the monitoring tasks between different switches in order to reduce the number of rules needed in each switch. This method is orthogonal to our distributed solution (see Section~\ref{Section:distributed}), and can be combined to further reduce the number of switch entries. 


Another recent work,~\cite{DREAM}, proposes DREAM, a framework for identifying heavy hitters (see Section~\ref{Section:FreqItems}) in traffic using TCAM based hardware. As shown in~\cite{DREAM}, the algorithm they use for heavy hitters detection may require more TCAM entries than a commodity switch may have available. Therefore DREAM performs efficient multi-switch resource allocation between switches to achieve the desired accuracy rates. The \algfull algorithm we propose (Section~\ref{Section:SampleAndPick}) requires significantly less counters in the switch and can be used by DREAM to reduce the overall number of switch entries used.


\section{Conclusions}
\label{Section:conclusions}

We have presented techniques for performing large flow detection and sampling in SDN. Our sampling techniques are unique in that they are simple and remain mostly within the confinements of the OpenFlow standard. Our approximation algorithms for large flows detection provide a generic mechanism for SDN, providing a way to detect various types of large flows with a relatively small error rate while minimizing the computation and space overhead in the switch and requiring little controller-switch communication. Furthermore, we expanded our algorithms to a distributed multi-switch setting.





\bibliographystyle{IEEEtran}
\bibliography{biblio}


\end{document}